\documentclass[a4paper,11pt]{article}

\usepackage{charter}   %% ÆÁÄ»ÔĶÁºÜºÃ¡£
\usepackage[charter]{mathdesign}

\usepackage{graphicx}
\usepackage{hyperref}
\usepackage{enumerate}
\usepackage[top=1.0cm,bottom=1.0cm,left=2.1cm,right=2.1cm,includehead,includefoot]{geometry}  
\usepackage{threeparttable}
\usepackage{booktabs}

\begin{document}
\title{A General Theory for Direct Quantitative Analysis of Antigen}
\author{Feng Gan$^{1}$ \thanks{Corresponding author, Email: cesgf@mail.sysu.edu.cn}, Gang Ye$^2$,  Langxia Liu$^3$, Xiaohui Liu$^3$\\
1. School of Chemistry and Chemical Engineering, Sun Yat-Sen University, \\Guangzhou 510275, P.R. China\\2. Guangzhou Oversees Chinese Hospital, Guangzhou 510630, China\\3. College of Life Science and Technology, Jinan University, Guangzhou 510630, China}

\maketitle

\begin{abstract}
A theory for direct quantitative analysis of an antigen is proposed. It is based on a potential homogenous immunoreaction system. It establishes an equation to describe the concentration change of the antigen and antibody complex. A maximum point is found in the concentration profile of the complex which can be used to calculate the concentration of the antigen. An experimental scheme was designed for a commercial time-resolved fluoroimmunoassay kit for HBsAg, which is based heterogeneous immunoreaction.  The results showed that the theory is practically applicable. 
\end{abstract}

\section{Introduction}

Antigen is a substance that was introduced into the body of  a living organism, which will lead to a disease sometimes. The invading of antigen will trigger the generation of an substance called antibody by the immune system of the organism. The antibody plays the role to kill or neutralize the antigen. This kind of reaction between antigen and antibody have found application in qualitative analysis and quantitative analysis of antigen.  The main topic of this paper focuses on the quantitative analysis.

A well known method to detect an antigen is the enzyme-linked immunosorbent assay (ELISA)\cite{Player1993p67-72, Lien2000p301-309, Lee2001p19-23}. It mostly focuses on detecting the presence of an antigen. With the increasing demand on quantitative analysis of antigen in the diagnosis and treatment of a diesease such as hepatitis B, the time-resolved fluoroimmunoassay (TRFIA) gradually shows its advantages\cite{Meurman1982p920-925,Siitari1983p258-260,Roberts1991p49-56} . The TRFIA makes the best use of fluorescence delay effect of europium (Eu) to eliminates the background fluorescence\cite{Degan1999p215-222}. So, the TRFIA is especially adapted in complicated sample such as sera.

There is a problem with the existing quantitative methods for antigens. They usually need the standard substance of the antigen in the calibration procedure of a measurement\cite{Rodella2006p206-212}. This procedure has been an indispensable part of presently existing quantitative analysis based on analytical instruments. However, the expense on the standard substance is a huge burden when the substance is expensive. Furthermore, when the standard substance is not available, the quantitative analysis can not be implemented. 

The purpose of this paper is to establish a general theory of quantitative analysis  that does not rely on the standard substance of target antigen.  The theory was developed based on  a potential homogenous immunoreaction system. It was finally applied to the direct quantitative analysis of HBsAg by designing an experimental scheme for heterogeneous immunoreaction.

\section{Theory}
Figure \ref{fig:1} shows the fundamental principal and experimental steps of a commerical time-resolved fluoroimmunoassy (TRFIA) kit for HBsAg. The double antibodies sandwich technology ensures that one antigen (Ag) combines with one labeled antibody (Abx) \cite{Mathis1997p3011-3014}. This kind of reaction mode is the requirement for quantiative analysis of the Ag. Thus, the reaction between Ag and Abx can be expressed as follows:
\begin{equation}
  \label{eq:1}
  \rm Ag + Abx ~~\rightarrow ~~ Ag\cdot Abx
\end{equation}
where x represents a label such as Eu$^{3+}$. Final measurement is usually based on the label. We assume the initial concentrations of Ag and Abx are $C_{\rm Ag}$ and $C_{\rm Abx}$, respectively. When $V_t$ volume of Abx's solution is added into $V_0$ volume of Ag's solution, the mass balances are as following
\begin{equation}
  \label{eq:2}
  \frac{C_{\rm Ag}V_t}{V_0 + V_t} = [C_{\rm Ag}] + [{\rm Ag\cdot\rm{Abx}}]
\end{equation}
\begin{equation}
  \label{eq:3}
  \frac{C_{\rm Abx}V_t}{V_0 + V_t} = [C_{\rm Abx}] + [{\rm Ag\cdot\rm{Abx}}]
\end{equation}
where [ ] represents the equilibrium concentration of the corresponding species. From equation (\ref{eq:2}) and equation (\ref{eq:3}), we get
\begin{equation}
  \label{eq:4}
  [{\rm Ag\cdot\rm{Abx}}] = \frac{{g(\rho ) - \sqrt {g^2 (\rho ) - h(\rho )} }}{2}
\end{equation}
where $ \rho  = \frac{{V_t }}{{V_0 }} $;$
g(\rho ) =  \frac{ \rho C_{\rm Abx} + C_{\rm Ag} }{{1 + \rho }} + \frac{1}{K}
$; $
h(\rho ) = \frac{{4\rho C_{\rm Ag} C_{\rm Abx} }}{{(1 + \rho )^2 }}
$;
$ K= \frac{[C_{\rm Ag\cdot\rm{Abx}}]}{[C_{\rm Ag}][C_{\rm Abx}]}$.

There is a maximum point in the plot of  $ [{\rm Ag\cdot {Abx}}]$ versus   $\rho$ according to equation (\ref{eq:4}). The first derivation of  $ [{\rm Ag\cdot {Abx}}]$ versus $\rho $ is
\begin{equation} \label{eq:5}
 [{\rm Ag\cdot Abx}]' = \displaystyle \frac{g'(\rho )}{2} - \displaystyle \frac{1}{4}[g^2(\rho ) - h(\rho )]^{-
\frac{1}{2}}[2g(\rho )g'(\rho ) - h'(\rho )]
\end{equation}
where $g'(\rho ) =  \frac{C_{\rm Abx} -C_{\rm Ag} }{(1 + \rho )^2}$; $h'(\rho ) =
\frac{4C_{\rm Ag} C_{\rm Abx} (1 - \rho )}{(\rho + 1)^3}$.

Let $ [{\rm Ag\cdot Abx}]' =0$, we finally get
\begin{equation}\label{eq:6}
\rho_{_0} = \frac{ C_{\rm Ag}  C_{\rm Abx} \pm \sqrt {\left( { C_{\rm Ag} + C_{\rm Abx} +
\frac{1}{K}} \right)\frac{(  C_{\rm Ag} - C_{\rm Abx} )^2}{K}} }{  C_{\rm Abx} ^2 +
\frac{ C_{\rm Abx} - C_{\rm Ag}  }{K}}
\end{equation}
where $\rho_{_0}$ is the $\rho$ value at the maximum point.

And it can be proved that the $\rho_{_0}$ at the maximum point will be 
\begin{equation}
  \label{eq:7}
\mathop {\lim }\limits_{K \to \infty } \rho_{_0}  = \frac{C_{\rm Ag} } {C_{\rm Abx} }.
\end{equation}

Equation (\ref{eq:7}) shows that the concentration of the Ag can be calculated without using calibration procedure. The requisite condition is that the generated antigen antibody complex, the $\rm Ag\cdot Abx$, is stable enough. The relative error of the estimation is
\begin{equation}
\label{eq:8} 
r.e. = \frac{\rho_{_0} -  \frac{C_{\rm Ag} }{C_{\rm Abx} }}{\frac{C_{\rm Ag}
}{C_{\rm Abx} }} = \frac{C_{\rm Abx}^2\pm \frac{C_{\rm Abx} }{C_{\rm Ag} }\sqrt
{\left( {C_{\rm Ag} + C_{\rm Abx} + \frac{1}{K}} \right)\frac{(C_{\rm Ag} - C_{\rm
Abx} )^2}{K}} }{C_{\rm Abx} ^2 + \frac{ C_{\rm Abx} - C_{\rm Ag} }{K}} - 1.
\end{equation}

Obviously, $\lim \limits_{K \to \infty } r.e. = 0$. Similar results can be obtained when $C_{\rm Ag} = C_{\rm Abx}$. 

Above theory is actually based on homogeneous system. The final measurement is based on labeled antiboy.  It needs quite an effort to apply the theory to practical system because of the lack of proper reagent kit. This situation could be changed in near future by using the homogeneous fluroimmunoassay (HFIA)\cite{Mathis1997p3011-3014}. However, a properly designed experimental scheme can apply the theory in another way.

\section{An application to TRFIA of HBsAg}
\begin{figure}[htpb]
  \centering
  \includegraphics[width=8cm]{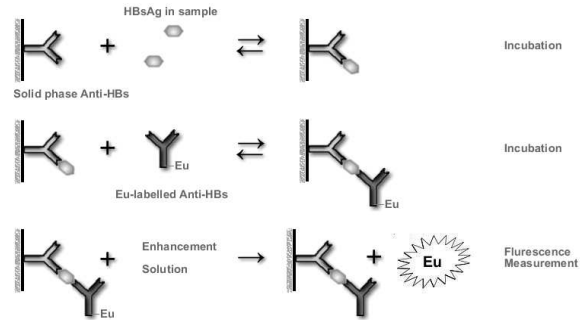}
  \caption{The basic steps of a commercial TRFIA kit for HBsAg. }
  \label{fig:1}
\end{figure}

Figure \ref{fig:1} shows a standard scheme for quantitative analysis of HBsAg using TRFIA. In the first incubation, the added HBsAg is fixed by the solid phase anti-HBs in the well. So, in the second incubation, the introduced Eu-labeled anti-HBs will combine with the HBsAg in 1:1 ratio. The measured fluorescence intensity of Eu in final step will be directly propotional to the quantity of the HBsAg. 

We designed an experimental scheme and shown it in Table \ref{tab:1}. In the scheme, the quantities of HBsAg in the wells are fixed which limits the binding quantity of the Eu-labelled anti-HBs in each of the wells. For the convenience of discussion, we assume the HBsAg quantity in each well is $C_{\rm HBsAg}V_0$ mol. When $C_{\rm anti-HBs-Eu}V_t > C_{\rm HBsAg}V_0$, extra added Eu-labelled anti-HBs will be washed out. On the other hand, when $C_{\rm anti-HBs-Eu}V_t < C_{\rm HBsAg}V_0$, the added Eu-labelled anti-HBs is totally combined with the HBsAg. The equations (\ref{eq:2}) and (\ref{eq:3}) hold under the latter situation.

\begin{center}
\begin{threeparttable}[h!]
  \caption{An experimental scheme for direct quantitative analysis of HBsAg}\label{tab:1}
\begin{tabular}{p{4cm}cccccccc}
  \toprule
Wells & 1 & 2 & 3 & 4 & 5 & 6 & 7 & 8 \\\hline
$V_0~(\mu \rm L)$  \tnote{a} & 100 & 100 & 100 & 100 & 100 & 100 & 100 & 100  \\
$V_t~(\mu \rm L)$ \tnote{b} & 20 & 40  & 60 & 80 &100
&120 &140 &160  \\
$(V_0 + V_t)~(\mu \rm L)$\tnote{c} & 120 & 140& 160& 180& 200& 220& 240 & 260 \\
  \bottomrule
\end{tabular}
\begin{tablenotes}
  \item [a] Volume of HBsAg.
    \item [b] Volume of anti-HBs-Eu.
      \item [c] Volume of enhancement solution.
\end{tablenotes}
\end{threeparttable}
\end{center}

The more important thing is that the final measurement is based on the quantity of Eu released by enhancement solution rather than antigen or antibody or antigen antibody complex. So, if we put the Eu in a volume variation environment like equations (\ref{eq:2}) and (\ref{eq:3}), equation (\ref{eq:4}) can be used to describe its concentration change. Finally, equation (\ref{eq:7}) can be used to calculate the concentration of HBsAg. The third line in Table \ref{tab:1} simulates equation (\ref{eq:3}) by adding Eu-labelled anti-HBs into the wells progressively. The most important operation is listed in the last line of Table \ref{tab:1}, which offers the final Eu a circumstance of volume variation. So, Eu has same concentration change as the antigen and antibody complex as shown in equation (\ref{eq:4}). 

\section{Experiments}

The TRFIA kit for HBsAg were from Guangzhou Darui Antibody Engineering And Technology Co. Ltd.. The sera for quality control were taken as unknown samples. The instrument was the Victor X5 Multilabel Plate Reader of PerkinElmer. TopPette Pipettors (Dragon Medical (Shanghai) Ltd.) were used to add solutions. The volumes of the reagents are shown in Table \ref{tab:1}. Experimental operations were done in accordance with the kit's manual.  Five repeat measurements were implemented.

\section{Results}

\begin{figure}[h!]
  \centering
  \includegraphics[width=8cm]{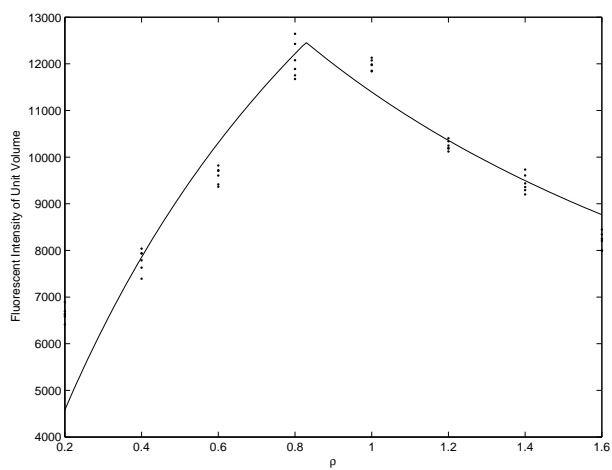}
  \caption{An example of fluorescence intensity of Eu with $\rho$.}
  \label{fig:2}
\end{figure}

Figure \ref{fig:2} shows the fluorescence intensities of Eu versus volume ratio. The solid line is the curve fitting result of the mean intensities using equation (\ref{eq:4}). The maximum point in the curve is used to calculate the concentration of the HBsAg. Table \ref{tab:2} shows the calculated results of two sera. It must note that fluorescence intensity should be adjust to unit volume because the instrument only recordes total fluorescence intensity.

\begin{table}[!h]
  \centering
  \caption{Measurement results of HBsAg ($C_{\rm anti-HBs-Eu}=0.00627\mu$mol/L)}
  \label{tab:2}
  \begin{tabular}{p{3.0cm}p{4cm}p{3.0cm}p{1.4cm}}\toprule
    Real concentration of HBsAg &	Mean $\rho$ value at max point, variance and repeat number&	Calculated concentration of HBsAg&	Relative error\\\hline
0.00417 $\mu$mol/L &	0.760,  0.035,  5&	0.00477 $\mu$mol/L&	14.3\%\\
0.00625 $\mu$mol/L&	0.813,  0.0050,  5&	0.00510 $\mu$mol/L&	-18.5\%\\\bottomrule
  \end{tabular}
\end{table}

\section{Conclusion}
We successfully established a general theory for the direct quantitative analysis of antigen. The theory is self-contained mathematically. The application examples also show practicability of the theory.   The advantage of the theory can be shown when the standard substances are extremely expensive or there are even no standard substances available.  More experiments are still needed to make it a routine mehtod.

\section*{Acknowledgments}
Fundings from the NSFC under grant number 20875106 and the GDSF under grant number 9151027501000003 are gratefully
acknowledged.

\end{document}